\documentstyle[12pt]{article}
\begin{document}
\title{\rightline {\normalsize {$
\begin{array}{c}
{\bf \ UG-FT-58/96 \ } \\ {\bf \ TP-USL/96/08 \ } \\ {\rm hep-ph-9603223}
\end{array}
$}}
\vskip .3cm 
CP phases in models with some fermion \\ masses vanishing
and/or degenerate \thanks{%
Dedicated to Prof. Wojciech Kr\'olikowski in honour of his 70th birthday} }
\author{{\normalsize Francisco del Aguila} \\ 
{\normalsize Departamento de F\'\i sica Te\'orica y del Cosmos}\\
{\normalsize Universidad de Granada, 18071 Spain} \\ 
{\normalsize Marek Zra{\l}ek } \\ {\normalsize Field Theory and Particle
Physics Department}\\ {\normalsize University of Silesia, 40-007 Katowice,
Poland}\\ 
\\ 
\centerline{\it (to appear in Acta Physica Polonica)}
}
\maketitle
\begin{abstract}
{\normalsize {We count the number of CP breaking phases in models with $%
SU(2)_L\times U(1)_Y$ and $SU(2)_L\times SU(2)_R\times U(1)_{B-L}$
electroweak gauge groups and extended matter contents with some fermion
masses vanishing and/or degenerate. Quarks and leptons, including Majorana
neutrinos, are treated in a similar way. CP violation is characterized in
the mass-eigenstate and in the weak-eigenstate bases. Necessary and
sufficient conditions for CP conservation, invariant under weak basis
redefinitions are also studied in these models. CP violating factors
entering in physical observables and only invariant under phase
redefinitions are discussed.} }
\end{abstract}
%
%
\newpage
\baselineskip16pt 
\section{Introduction}
CP violation is related to the presence of complex phases in the mixing
matrices describing the gauge couplings in the mass-eigenstate basis. (We do
not consider other (Higgs) sources of CP violation.) However, not all phases
in the mixing matrices are CP violating. Some of them can be eliminated
redefining the fermion phases. In the standard model with three
non-degenerate quark families the six phases defining the $3\times 3$
unitary mixing matrix reduce to one after an appropriate fermion field phase
redefinition. This was first realized by Kobayashi and Maskawa \cite{KM} and
it is the simplest way to account for the observed CP violation \cite{PDB}.
In general if there are degenerate fermion masses, the number of CP
violating phases is further reduced. In the standard model with three
massless neutrinos the six phases defining the $3\times 3$ unitary mixing
matrix in the lepton sector can be eliminated. As a matter of fact not only
all the phases but the three real mixing angles are non-physical. There is
no mixing between lepton families and the three lepton numbers are
conserved. These two cases are extreme, non-degenerate quark masses and
degenerate and vanishing neutrino masses. Here we will examine the
intermediate case. We allow for an arbitrary number of standard families and
Majorana neutrinos with some fermion masses vanishing and/or degenerate. We
consider models with the standard gauge group $SU(3)_C \times SU(2)_L\times
U(1)_Y$ and with its left-right extension $SU(3)_C\times SU(2)_L\times
SU(2)_R\times U(1)_Y$, in turn.
Although the number of CP violating phases is more easily counted in the
mass-eigenstate basis, CP violation can be also discussed in the
weak-eigenstate basis \cite{S,V}. In this basis the mass matrices are in
general non-diagonal and necessarily complex if CP is not conserved. The use
of an invariant formulation for CP conservation is more convenient in this
case. Necessary (and sufficient) conditions for CP conservation can be found
which are independent of the choice of basis \cite{J,BBG,BR,AZ}. If one of
these conditions is not fulfilled, CP is not conserved. However, some of
them could be trivially satisfied if some fermion masses are vanishing or
degenerate. We study this possibility in simple cases.
In Section 2 we count the number of CP breaking phases in $SU(3)_C\times
SU(2)_L\times U(1)_Y$ gauge models with some vanishing and/or degenerate
quark and lepton masses, including Majorana neutrinos. $SU(3)_C\times
SU(2)_L\times SU(2)_R\times U(1)_Y$ gauge models are considered in Section
3. The necessary and sufficient conditions for CP conservation are discussed
in Section 4 in simple models with some vanishing and/or degenerate fermion
masses. In Section 5 we comment on the relevance of the invariants under
fermion phase redefinitions. Section 6 is devoted to conclusions.
\section{The standard model with $n_L$ fermion families and $n_R$ neutral
fermion singlets}
The $n_L$ left-handed fields transform as $SU(2)_L$ doublets, whereas the $%
n_R$ right-handed fields are $SU(2)_L$ singlets. Hence, only the left-handed
fermions interact with the charged gauge boson $W$.
\subsection{The quark sector}
Let $M_u$ and $M_d$ be the $n_L\times n_L$ mass matrices for up and down
quarks, respectively, in the weak-eigenstate basis ($n_L = n_R$). In general
they are complex and can be diagonalized by unitary transformations 
\begin{equation}
\left( M_u \right) _{diag} = U_L^{u\dagger} M_u U_R^u, \ \ \left( M_d
\right) _{diag} = U_L^{d\dagger} M_d U_R^d, 
\end{equation}
where $U_{L,R}^{u,d}$ are $n_L\times n_L$ unitary matrices. Thus, the
Cabibbo-Kobayashi-Maskawa (CKM) mixing matrix \cite{KM} reads 
\begin{equation}
U_{CKM} = U_L^{u\dagger} U_L^d. 
\end{equation}
The matrices $U_{L,R}^{u,d}$ in Eq. (1) are not uniquely determined. We can
still perform unitary transformations $V_{L,R}^{u,d}$, which leave unchanged
the diagonal mass matrices 
\begin{equation}
\left( M_u\right) _{diag}=V_L^{u\dagger} \left( M_u\right)_{diag} V_R^u, \
\left( M_d\right) _{diag}=V_L^{d\dagger} \left( M_d\right)_{diag} V_R^d, 
\end{equation}
but redefine the CKM matrix 
\begin{equation}
U_{CKM}\rightarrow V_L^{u\dagger} U_{CKM} V_L^d = \left( U_L^u V_L^u
\right)^{\dagger} \left( U_L^d V_L^d \right). 
\end{equation}
How the matrices $V_{L,R}^u$ ($V_{L,R}^d$) look like depends on the
properties of $\left( M_u\right) _{diag}$ ($\left( M_d\right) _{diag}$). Let
us assume that $\left( M_u\right) _{diag}$ ($\left( M_d\right) _{diag})$ has 
$l_u^0$ ($l_d^0$) vanishing masses, $l_u$ ($l_d$) $> 1$ degenerate masses $%
m_u (m_d)$, and $n_L-l_u^0-l_u$ ($n_L-l_d^0-l_d$) non-degenerate masses 
\begin{equation}
\left( M_u\right) _{diag}=\left( 
\begin{array}{ccc}
\begin{array}{ccc}
0 &  &  \\  
& \ddots &  \\  
&  & 0 
\end{array}
&  & 0 \\  
& 
\begin{array}{ccc}
m_u &  &  \\  
& \ddots &  \\  
&  & m_u 
\end{array}
&  \\ 
0 &  & 
\begin{array}{ccc}
m_1 &  &  \\  
& m_2 &  \\  
&  & \ddots 
\end{array}
\end{array}
\right) 
\begin{array}{c}
\left. 
\begin{array}{c}
\\  
\\  
\\  
\end{array}
\right\} l_u^0 \ \ \ \ \ \ \ \ \ \ \ \  \\ 
\left. 
\begin{array}{c}
\\  
\\  
\\  
\end{array}
\right\} l_u \ \ \ \ \ \ \ \ \ \ \ \  \\ 
\left. 
\begin{array}{c}
\\  
\\  
\\  
\end{array}
\ \right\} n_L-l_u^0-l_u 
\end{array}
\end{equation}
(analogously for $(M_d)_{diag}$). Then 
\begin{equation}
\begin{array}{c}
V_L^u=\left( 
\begin{array}{ccccc}
W_{uL}^0 &  &  &  & 0 \\  
& W_{uL} &  &  &  \\  
&  & e^{i\delta _1} &  &  \\  
&  &  & \ddots &  \\ 
0 &  &  &  & e^{i\delta _{^{n_L-l_u^0-l_u}}} 
\end{array}
\right) , \\  
\\ 
V_R^u=\left( 
\begin{array}{ccccc}
W_{uR}^0 &  &  &  & 0 \\  
& W_{uL} &  &  &  \\  
&  & e^{i\delta _1} &  &  \\  
&  &  & \ddots &  \\ 
0 &  &  &  & e^{i\delta _{n_L-l_u^0-l_u}} 
\end{array}
\right) , 
\end{array}
\end{equation}
where $W_{uL,R}^0$ ($W_{uL}$) are $l_u^0\times l_u^0$ ($l_u\times l_u$)
unitary matrices, and analogously for $V^d_{L,R}$.
Now we can count the number of CP violating phases. The CKM matrix is a $%
n_L\times n_L$ unitary matrix and is parametrized by $\frac{n_L\left(
n_L-1\right) }2$ mixing angles and $\frac{n_L\left( n_L+1\right) }2$ complex
phases. Not all of these parameters are physical but we can get rid of the
unphysical ones as shown in Eq. (4) with an appropriate choice of $V^{u,d}_L$
in Eq. (6). Although known results [3] can be easily recovered as we do
below, the general case is involved. It looks necessary to treat it with a
computer \cite{AAZ}.
\begin{itemize}
\item  {2.1.1.} The counting for $l_{u,d}^0=0,1,\ l_{u,d}=0$ is the same as
in the non-vanishing, non-degenerate standard case because $V_L^{u,d}$ have
the same structure. The number of CP violating phases is equal to the number
of phases in a $n_L\times n_L$ unitary matrix, $\frac{n_L\left( n_L+1\right) 
}2$, minus the number of phases in $V_L^{u,d}$, $2n_L$, plus 1 to avoid
double counting of the common phase redefinition 
\begin{equation}
\frac{n_L\left( n_L+1\right) }2-2n_L+1=\frac{\left( n_L-1\right) \left(
n_L-2\right) }2.
\end{equation}
For $n_L=3$ one recovers the standard model result with one CP violating
phase.
\item  {2.1.2.} For $l_u^0$ or $l_u=n_L-1$ there is no CP violation. Using
the decomposition of a $n_L\times n_L$ unitary matrix 
\begin{equation}
\begin{array}{cc}
U_{CKM}=\left( 
\begin{array}{cc}
W_{uL}^0 & 0 \\ 
0 & 1
\end{array}
\right)  &  \\  
&  \\ 
\times \left( 
\begin{array}{cccccc}
c_1 & -s_1s_2 & -s_1c_2s_3 & \dots  & \dots  & -s_1c_2\dots c_{n_L-1} \\  
& c_2 & -s_2s_3 & \dots  & \dots  & -s_2c_3..c_{n_L-1} \\  
&  & c_3 & \dots  & \dots  & -s_3c_4.c_{n_L-1} \\  
&  &  & \dots  & \dots  & \dots  \\ 
0 &  &  &  & c_{n_L-1} & -s_{n_L-1} \\ 
s_1 & c_1s_2 & c_1c_2s_3 & \dots  & \dots  & c_1c_2\dots c_{n_L-1}
\end{array}
\right)  &  \\  
&  \\ 
\times \left( 
\begin{array}{ccc}
e^{-i\delta _1} &  &  \\  
& \ddots  &  \\  
&  & e^{-i\delta _{^{n_L}}}
\end{array}
\right) , & 
\end{array}
\end{equation}
where $W_{uL}^0$ is a $(n_L-1)\times (n_L-1)$ unitary matrix, and Eqs. (4,6) 
$U_{CKM}$ can be made always real, almost triangular and depending only on $%
n_L-1$ mixing angles. For $l_d^0$ or $l_d=n_L-1$ we use the inverse unitary
decomposition to prove that there is no CP violation either.
\end{itemize}
\subsection{The lepton sector}
$n_L$ is the number of standard fermion families and $n_R$ the number of
right-handed neutral fermion singlets. Then the charged lepton mass matrix $%
M_l$ is $n_L\times n_L$ and complex, and the neutrino mass matrix $M_\nu $
is $\left( n_L+n_R\right) \times \left( n_L+n_R\right) $, complex and
symmetric. They can be diagonalized by unitary transformations 
\begin{equation}
\left( M_l\right) _{diag}=U_L^{l\dagger }M_lU_R^l,\ \left( M_\nu \right)
_{diag}=U^TM_\nu U,
\end{equation}
with $U_{L,R}^l$ and $U$ $n_L\times n_L$ and $(n_L+n_R)\times (n_L+n_R)$
unitary matrices, respectively. Defining 
\begin{equation}
\begin{array}{ccc}
& \overbrace{}^{n_L+n_R} &  \\ 
U= & \left( 
\begin{array}{c}
U_L^{*} \\ 
U_R
\end{array}
\right)  & 
\begin{array}{c}
\}n_L \\ 
\}n_R
\end{array}
,
\end{array}
\end{equation}
the mixing matrices in the charged and neutral currents can be written 
\begin{equation}
K=U_L^{\dagger }U_L^l\ {\rm and}\ \Omega =KK^{\dagger },
\end{equation}
respectively. The diagonalization conditions in Eq. (9) do not determine $%
U_{L,R}^l$ and $U$ uniquely. One can still perform unitary transformations
which leave unchanged the (diagonal) mass matrices 
\begin{equation}
\begin{array}{c}
\left( M_l\right) _{diag}=V_L^{\dagger }\left( M_l\right) _{diag}V_R=\left(
U_L^lV_L\right) ^{\dagger }M_l\left( U_R^lV_R\right) , \\ 
\left( M_\nu \right) _{diag}=V^T\left( M_\nu \right) _{diag}V=\left(
UV\right) ^TM_\nu \left( UV\right) .
\end{array}
\end{equation}
The form of $V_{L,R}$ and $V$ depends on the fermion spectrum (degeneracy).
For $l_l^0$ vanishing, $l_l$ degenerate and $n_L-l_l^0-l_l$ non-degenerate
charged lepton masses and $l_\nu ^0$ vanishing, $l_\nu $ degenerate and $%
n_L+n_R-l_\nu ^0-l_\nu $ non-degenerate neutrino masses 
\begin{equation}
\begin{array}{c}
V_L=\left( 
\begin{array}{ccccc}
W_L^0 &  &  &  & 0 \\  
& W_L &  &  &  \\  
&  & e^{i\alpha _1} &  &  \\  
&  &  & \ddots  &  \\ 
0 &  &  &  & e^{i\alpha _{n_L-l_l^0-l_l}}
\end{array}
\right) , \\  
\\ 
V_R=\left( 
\begin{array}{ccccc}
W_R^0 &  &  &  & 0 \\  
& W_L &  &  &  \\  
&  & e^{i\alpha _1} &  &  \\  
&  &  & \ddots  &  \\ 
0 &  &  &  & e^{i\alpha _{n_L-l_l^0-l_l}}
\end{array}
\right) , \\  
\\ 
V=\left( 
\begin{array}{cccccc}
W^0 &  &  &  &  & 0 \\  
& W &  &  &  &  \\  
&  & \pm 1 &  &  &  \\  
&  &  & \pm 1 &  &  \\  
&  &  &  & \ddots  &  \\ 
0 &  &  &  &  & \pm 1
\end{array}
\right) ,
\end{array}
\end{equation}
with $W_{L,R}^0,W_L,W^0$ and $W$ $l_l^0\times l_l^0,l_l\times l_l,l_\nu
^0\times l_\nu ^0$ unitary and $l_\nu \times l_\nu $ real orthogonal
matrices, respectively. The mixing matrices, however, do change under these
transformations 
\begin{equation}
K\rightarrow V^TKV_L,\ \Omega \rightarrow V^T\Omega V^{*}.
\end{equation}
The counting of CP violating phases in the lepton sector reduces to count
the phases in $K$ because the phases in $\Omega $ are not independent (see
Eq. (11)). $K$ is defined by the first $n_L$ columns of a $(n_L+n_R)\times
(n_L+n_R)$ unitary matrix. Thus it is parametrized by $\frac {n_L\left(
n_L+2n_R-1\right)}{2}$ mixing angles and $\frac {n_L\left(
n_L+2n_R+1\right)}{2}$ phases. But not all of them are physical. Eq. (14)
allows to subtract the unphysical ones. The general counting is involved but
it can be worked out with a computer \cite{AAZ}.
\begin{itemize}
\item  {2.2.1.} If there are no vanishing or degenerate lepton masses, the
number of CP breaking phases is equal to the number of phases parametrizing
$K$ minus the $n_L$ phases fixing $V_L$ [4,8], 
\begin{equation}
\frac{n_L\left( n_L+2n_R+1\right) }2-n_L
=\frac{n_L\left( n_L+2n_R-1\right) }2.
\end{equation}
If there is one massless neutrino, $W^0$ is one-dimensional and there is one
phase less.
\item  {2.2.2.} If, for instance, $n_L=n_R$ and there are $n_L$ massless
neutrinos, the number of CP violating phases in Eq. (15) is reduced by $%
\frac{n_L\left( n_L+1\right) }2$, which is the number of phases of the
unitary matrix $W^0$ in $V$. The subtraction of non-physical phases is more
delicate when the $n_L$ neutrinos have a common mass. The real and imaginary
parts of $K$ rotate independently because $W\subset V$ is a real orthogonal
matrix in this case. Then with an appropriate choice of $W$ $\frac{n_L\left(
n_L-1\right) }2$ entries in the ($n_L\times n_L$) $K$ upper half can be made
real, reducing the number of phases by the same amount. The unitarity
constraints and the charged lepton phase redefinitions can be chosen to fix
part of the phases in the $K$ lower half and are already subtracted in Eq.
(15). If the $n_L$ massless or degenerate leptons are the charged ones, the
number of CP violating phases in Eq. (15) is reduced by $\frac{n_L\left(
n_L-1\right) }2$, which is the number of phases in $W_L^0$ or $W_L$ in $V_L$%
, $\frac{n_L\left( n_L+1\right) }2$, minus the number of diagonal
phases already subtracted in Eq. (15), $n_L$.
\item  {2.2.3.} If $n_L=n_R=1$ and both neutrinos have a common mass, CP is
conserved. $K$, which is $2\times 1$, depends on 2 complex numbers. With an
appropriate choice of the $2\times 2$ real orthogonal matrix $W=V$ the
moduli of $K_{11}$ and $K_{21}$ can be made equal. Then redefining the
charged lepton phase we can always assume $K=\left( 
\begin{array}{c}
a \\ 
a^{*}
\end{array}
\right) $. The phase of $a$, however, does not stand for CP violation. The
lagrangian is invariant under complex conjugation and the interchange of
both neutrinos. Similarly for $n_L=2,n_R=0$, with an appropriate choice of $W
$ the moduli of the $2\times 2$ unitary matrix $K$ can be made equal. Then
redefining the charged lepton phases we can always assume $K=\left( 
\begin{array}{cc}
a & a \\ 
-a^{*} & a^{*}
\end{array}
\right) $. The phase of $a$ does not stand for CP violation either. The
lagrangian is invariant under complex conjugation, the interchange of both
neutrinos and the change of sign of the charged lepton in the first column.
If there are the two charged leptons which are degenerate (or massless), CP
is also conserved because in this case $V_L$ in Eq. (14) is an arbitrary $%
2\times 2$ unitary matrix.
\end{itemize}
\section{Left-right models with $n$ fermion families}
In this case $n_L = n_R = n$ and there are left-handed as well as
right-handed charged currents.
\subsection{The quark sector}
Without any additional symmetry M$_{u}$ and M$_d$ are arbitrary complex
matrices and the expressions in Section 2 are still valid. However, in
addition to the CKM matrix for left-handed currents (see Eq. (2)) 
\begin{equation}
U_{CKM}^L = U_L^{u\dagger} U_L^d, 
\end{equation}
there is a CKM mixing matrix for right-handed currents 
\begin{equation}
U_{CKM}^R = U_R^{u\dagger} U_R^d. 
\end{equation}
Both $n\times n$ unitary matrices can be redefined without modifying the
(diagonal) mass matrices (see Eqs. (3-6)) 
\begin{equation}
U_{CKM}^L\rightarrow V_L^{u\dagger} U_{CKM}^L V_L^d, \ U_{CKM}^R\rightarrow
V_R^{u\dagger} U_{CKM}^R V_R^d. 
\end{equation}
The counting of CP breaking phases in $U_{CKM}^L$ is the same as the
standard model counting in Section 2. On the other hand, the number of CP
breaking phases in $U_{CKM}^R$ can be reduced by an appropriate choice of $%
W^0_{u,d\ R}$ (see Eq. (6)), for the other entries in $V_R^{u,d}$ are fixed
by the corresponding entries in $V_L^{u,d}$. However, one may choose to
eliminate the non-physical phases in $U_{CKM}^{L,R}$ in a different way.
What matters it is the combined number of CP violating phases.
\begin{itemize}
\item  {3.1.1.} CP can be violated in a left-right model even for one
family, $n=1$, if both quarks are massive. This is so because the phase
redefinition in Eq. (18) is the same for $U_{CKM}^L$ and $U_{CKM}^R$, as
there are the same $V_L^u(V_L^d)$ and $V_R^u(V_R^d)$. Thus the number of CP
violating phases is $2-1=1$. Similarly for three generations, $n=3$, CP can
be broken if there is at least one massive quark of each type.
\end{itemize}
\subsection{The lepton sector}
In this sector there are also right-handed currents and there are left- and
right-handed mixing matrices (see Eq. (11)) 
\begin{equation}
\begin{array}{ccc}
& K_L = U_L^{\dagger} U_L^l, & K_R = U_R^{\dagger} U_R^l \\ 
{\rm and} &  &  \\  
& \Omega _L = K_L K_L^{\dagger}, & \Omega _R = K_R K_R^{\dagger} . 
\end{array}
\end{equation}
The $K$ matrices are $2n\times n$ and satisfy the orthogonality condition 
\begin{equation}
K_L^T K_R = 0; 
\end{equation}
and the $\Omega $ matrices, which are $2n\times 2n$, are completely fixed by
the $K$ matrices. As in Eq. (14) the mass matrices remain unchanged, whereas 
\begin{equation}
K_L\rightarrow V^T K_L V_L, \ K_R\rightarrow V^{\dagger} K_R V_R. 
\end{equation}
The unphysical phases in $K_L$ can be eliminated as in Section 2. The phases
in $K_R$ can be also eliminated using Eq. (21) but only if there is any
freedom left after fixing $V$ and $V_R$, which is related to $V_L$ (Eq.
(13)), to reduce the number of $K_L$ phases. What matters is the combined
number of CP violating phases in $K_L$ and $K_R$.
\begin{itemize}
\item  {3.2.1.} For $n=1$ there are in general 2 CP breaking phases, of the
four phases in $K_L$ and $K_R$ one is fixed by Eq. (20) and another one is
eliminated by an appropriate choice of $V_L$ ($V_R=V_L$ if the charged
lepton has a non-zero mass.) If the charged lepton is massless, $V_R$ is
independent of $V_L$ and we can get rid of a third phase. Finally, if one
neutrino is also massless, the fourth phase can be cancelled and CP is 
conserved. If the two neutrinos have a common mass, we can always write 
$K_L=\left( 
\begin{array}{c}
a \\ 
-a^{*}
\end{array}
\right) $, $K_R=\left( 
\begin{array}{c}
a^{*} \\ 
a
\end{array}
\right) e^{i\beta }$. 
The $\beta $ phase can be also eliminated if the charged
lepton is massless $\left( V_L\neq V_R\right) .$ The resulting form of 
$K_{L,R}$ is the same as in case 2.2.3 and CP is also conserved because the 
lagrangian is invariant under the same operations.
\item  {3.2.2.} For $n>1$ CP can be violated even in the case of $n$
degenerate charged leptons and $n$ massless plus $n$ degenerate heavy
neutrinos.
\end{itemize}
\section{CP symmetry breaking in the weak basis}
In the two previous Sections we have discussed CP symmetry breaking in the
mass-eigenstate basis. CP violation can be also studied in the weak basis
where gauge interactions are diagonal. CP conservation is then related to
the specific form of quark and lepton mass matrices, $M_u,M_d,M_l,M_\nu $ 
\cite{J,BBG,BR,AZ}. If these are real, CP is conserved. However, they can be
complex and CP be still conserved. This is so because we can perform unitary
transformations on the fields which leave unchanged the gauge couplings but
redefine the mass matrices. If there are only left-handed currents, these
transformations on quarks and leptons read 
\begin{equation}
\begin{array}{ccc}
u_L & \rightarrow  & X_Lu_L, \\ 
d_L & \rightarrow  & X_Ld_L, \\ 
u_R & \rightarrow  & X_R^uu_R, \\ 
d_R & \rightarrow  & X_R^dd_R, \\  
&  &  \\ 
\nu _L & \rightarrow  & Y_L\nu _L, \\ 
l_L & \rightarrow  & Y_Ll_L, \\ 
l_R & \rightarrow  & Y_R^ll_R, \\ 
\nu _R & \rightarrow  & Y_R^\nu \nu _R,
\end{array}
\end{equation}
where $X_L,X_R^u,X_R^d,Y_L$ and $Y_R^l$ ($Y_R^\nu $) are arbitrary $%
n_L\times n_L$ ($n_R\times n_R$) unitary matrices. If there are also
right-handed currents, 
\begin{equation}
X_R^u=X_R^d,\ Y_R^l=Y_R^\nu .
\end{equation}
Under these transformations the mass matrices $M_u,M_d,M_l$ and 
\begin{equation}
\begin{array}{cccc}
& \overbrace{}^{n_L} & \overbrace{}^{n_R} &  \\ 
M_\nu = & \left( 
\begin{array}{c}
M_L \\ 
M_D^T
\end{array}
\right.  & \left. 
\begin{array}{c}
M_D \\ 
M_R
\end{array}
\right)  & 
\begin{array}{c}
\}n_L \\ 
\}n_R
\end{array}
\end{array}
\end{equation}
change 
\begin{equation}
\begin{array}{ccc}
M_u & \rightarrow  & X_LM_uX_R^{u\dagger }, \\ 
M_d & \rightarrow  & X_LM_dX_R^{d\dagger }, \\  
&  &  \\ 
M_l & \rightarrow  & Y_LM_lY_R^{l\dagger }, \\ 
M_L & \rightarrow  & Y_LM_LY_L^T, \\ 
M_D & \rightarrow  & Y_LM_DY_R^{\nu \dagger }, \\ 
M_R & \rightarrow  & Y_R^{\nu *}M_RY_R^{\nu \dagger }.
\end{array}
\end{equation}
Then CP is conserved if and only if there exist unitary matrices $%
X_L,X_R^{u,d},Y_L,Y_R^{l,\nu }$ such that 
\begin{equation}
\begin{array}{ccc}
X_LM_uX_R^{u\dagger } & = & M_u^{*}, \\ 
X_LM_dX_R^{d\dagger } & = & M_d^{*}, \\  
&  &  \\ 
Y_LM_lY_R^{l\dagger } & = & M_l^{*}, \\ 
Y_LM_LY_L^T & = & M_L^{*}, \\ 
Y_LM_DY_R^{\nu \dagger } & = & M_D^{*}, \\ 
Y_R^{\nu *}M_RY_R^{\nu \dagger } & = & M_R^{*}.
\end{array}
\end{equation}
These conditions also suggest how to find other necessary (and sufficient)
CP invariant constraints which are more useful in practice
. In this Section we discuss these constraints when some fermion masses are
vanishing and/or degenerate (see Section 2,3).
\begin{itemize}
\item  {4.1.} In the standard model with $n_L=3$ generations of quarks the
necessary and sufficient condition for CP conservation is \cite{J,BBG} 
\begin{equation}
\begin{array}{c}
Tr\left[ M_uM_u^{\dagger },M_dM_d^{\dagger }\right] ^3=3Det\left[
M_uM_u^{\dagger },M_dM_d^{\dagger }\right] = \\ 
-6i\left( m_t^2-m_c^2\right) \left( m_t^2-m_u^2\right) \left(
m_c^2-m_u^2\right) \left( m_b^2-m_s^2\right) \left( m_b^2-m_d^2\right)  \\ 
\times \left( m_s^2-m_d^2\right) Im\left(
U_{ud}U_{cs}U_{us}^{*}U_{cd}^{*}\right) =0,
\end{array}
\end{equation}
where $m_f$ are the quark masses and $U$ is the CKM matrix. CP can be
violated if there is at most one massless quark of each type (case 2.1.1).
However, this invariant is identically zero if two up or down quark masses
are degenerate (case 2.1.2).
\item  {4.2.} For leptons practical, necessary and sufficient CP invariant
constraints were obtained in Ref. \cite{AZ} in simple cases . For $n_L=n_R=1$
the constraint for CP conservation is 
\begin{equation}
\begin{array}{c}
ImTr\left( M_D^{\dagger }M_LM_D^{*}M_R\right) = \\ 
m_1m_2\left( m_2^2-m_1^2\right) Im\left( K_{11}^2K_{21}^{*2}\right) =0,
\end{array}
\end{equation}
where $m_i$ are the neutrino masses and $K$ is the mixing matrix in Eq.
(11). If there is one massless neutrino or both neutrinos are degenerate, CP
is conserved (cases 2.2.1 and 2.2.3, respectively). For $n_L=2,n_R=0$ the CP
invariant constraint is 
\begin{equation}
\begin{array}{c}
ImDet\left( M_LM_l^{*}M_l^TM_L^{\dagger }-M_LM_L^{\dagger }M_lM_l^{\dagger
}\right) = \\ 
m_1m_2\left( m_2^2-m_1^2\right) (m_\mu ^2-m_e^2)^2Im\left(
K_{11}^2K_{21}^{*2}\right) =0,
\end{array}
\end{equation}
where $m_{1,2(e,\mu )}$ are the neutrino (charged lepton) masses and $K$ is
the mixing matrix. CP is conserved if there is a massless neutrino (case
2.2.1) or the neutrinos (charged leptons) are degenerate (case 2.2.3).
\item  {4.3.} In left-right models with $n=1$ there are two necessary and
sufficient invariant constraints for CP conservation in the lepton sector 
\begin{equation}
\begin{array}{c}
ImTr\left( M_lM_D^{\dagger }\right) =m_eIm\left(
m_1K_{L11}K_{R11}^{*}+m_2K_{L21}K_{R21}^{*}\right) =0, \\  
\\ 
ImTr\left( M_lM_R^{\dagger }M_l^TM_L^{\dagger }-M_LM_D^{*}M_RM_D^{\dagger
}\right) = \\ 
m_e^2Im\left(
(m_1K_{L11}^2+m_2K_{L21}^2)(m_1K_{R11}^{*2}+m_2K_{R21}^{*2})\right)  \\ 
-m_1m_2\left( m_2^2-m_1^2\right) Im(K_{L11}^{*2}K_{L21}^2)=0,
\end{array}
\end{equation}
where $m_{1,2(e)}$ are the neutrino (charged lepton) masses and $K_{L,R}$
are the mixing matrices. CP is conserved if the charged lepton is massless
and there is one massless neutrino or both neutrinos are degenerate (case
3.2.1). Otherwise, CP can be broken.
\end{itemize}
As in the former examples we expect that the counting of CP breaking phases
in Sections 2,3 will be useful for searching for a set of necessary and
sufficient, and also practical, CP invariant constraints.
\section{Phase redefinition invariants}
Physical observables can not depend on fermion field redefinitions. In the
mass-eigenstate basis the only fermion field redefinitions left are the
unitary transformations in Eqs. (6,13), which leave unchanged the diagonal
mass matrices and redefine the mixing matrices (see Eqs. (4,14,18,21)). Then
the corresponding observables can only depend on quantities invariant under
these transformations. The simplest of these quantities are 
\begin{equation}
\sum\limits_{i,j}\left| T_{ij}\right| ^2,
\end{equation}
where if $i$ or $j$ stands for a degenerate fermion, the sum (as the sums
below) also includes the other fermions of the same type with the same mass.
Otherwise, $i$ and $j$ can be any set of fermions and $T$ is any mixing
matrix, $U,K,\Omega $. However, these expressions do not depend on any
phase, and even if CP is conserved, they are in general non-zero. (Sums 
\begin{equation}
\sum\limits_jT_{ij}T_{kj}^{*}
\end{equation}
with $i\neq j$ are not invariant because even for non-degenerate Majorana
neutrinos they can transform with a sign (see Eq. (13)).) In left-right
models there are also mixed bilinear invariants 
\begin{equation}
\sum\limits_{i,j}T_{Lij}T_{Rij}^{*},
\end{equation}
where $T_L$ is a left-handed mixing matrix, $U_L,K_L,\Omega _L$, and $T_R$
its right-handed partner, $U_R,K_R,\Omega _R$. These invariants depend in
general on the CP breaking phases. The number of independent CP violating
invariants is, of course, finite in specific models. In the left-right model
in Section 4 with $n=1$ there are two such invariants (see Eq. (30)) 
\begin{equation}
K_{L11}K_{R11}^{*}\ {\rm and}\ K_{L21}K_{R21}^{*}.
\end{equation}
A non-zero imaginary part of these invariants stands for CP
non-conservation. In models with only left-handed mixing matrices we have to
look for invariants of higher dimensions to observe CP violation. Possible
invariants of dimension four are 
\begin{equation}
\sum\limits_{i,j,k,m}T_{ij}T_{km}T_{im}^{*}T_{kj}^{*}.
\end{equation}
In the minimal standard model CP violation is characterized by a non-zero
imaginary part of one of these invariants, for example (see Eq. (27)) $%
Im\left( U_{ud}U_{cs}U_{us}^{*}U_{cd}^{*}\right) $.
\section{Conclusions}
We have discussed the number of independent CP breaking phases in models
with $SU(2)_L\times U(1)_Y$ and $SU(2)_L\times SU(2)_R\times U(1)_{B-L}$
electroweak gauge groups and extended matter contents, paying special
attention to the case of vanishing and/or degenerate fermion masses. Quarks
and leptons, including Majorana neutrinos, are treated in a similar way. We
have also revised the necessary and sufficient constraints for CP
conservation in some simple models. Some of these constraints are
identically zero when some fermion masses vanish or are degenerate. The
knowledge of the number of independent CP violating phases and the study of
these particular cases are a useful guide for the search of CP invariant
constraints which are not only necessary but sufficient for CP conservation.
Observables involving well-defined mass eigenstates depend on factors which
are only invariant under phase redefinitions. We study the phase
redefinition invariants of lowest dimension.
\vskip 1.cm \noindent
{\bf Acknowledgements}
\vskip .5cm This work was partially supported by the CICYT under contract
AEN94-0936, the European Union under contract CHRX-CT92-0004, the Curie-Sk{%
\l }odowska grant MEN/NSF 93-145, and the Polish Committee for Scientific
Researches under grants PB 659/P03/95/08 and 2P30225206/93. We thank J.A.
Aguilar-Saavedra for discussions.

\end{document}